\documentclass[12pt]{article}
\usepackage{color}
\usepackage{graphicx}

\definecolor{violet}{rgb}{0.4,0,0.6}
\definecolor{vert}{rgb}{0,0.6,0.2}
\definecolor{navy}{rgb}{0.0,0.0,0.4}

\def\colorange#1{\textcolor[named]{Orange}{#1}}
\def\colbrun#1{\textcolor[named]{Brown}{#1}}

\def\spose#1{\hbox to 0pt{#1\hss}}\def\lta{\mathrel{\spose{\lower 3pt\hbox
{$\mathchar"218$}}\raise 2.0pt\hbox{$\mathchar"13C$}}}  \def\gta{\mathrel
{\spose{\lower 3pt\hbox{$\mathchar"218$}}\raise 2.0pt\hbox{$\mathchar"13E$}}} 
\def\Libra{\spose {--} {\cal L}}

\font\sixrm=cmr6

\def\sigme{{\color{vert}\sigma}} \def\xe{{\color{vert}\xi}} 
\def\ete{{\color{vert}\eta}} 
\def\delte{{\color{vert}\delta}}

\def\dP{\spose {\lower 2.0pt\hbox{$_{_\Gamma}$} } {\delte}}
\def\dL{\spose {\lower 5.0pt\hbox{\sixrm \color{vert}L} } {\delte}}
\def\dE{\spose {\lower 5.0pt\hbox{\sixrm E} } {\delta}}

\def\xiI{\spose {\raise 3.0pt\hbox{$\, \acute{\ }$}} {\xe}}
\def\xiJ{\spose {\raise 3.0pt\hbox{$\, \grave{\ }$}} {\xe}}
\def\deI{\spose {\raise 3.0pt\hbox{$\, \acute{\ }$}} {\delte}}
\def\deJ{\spose {\raise 3.0pt\hbox{$\, \grave{\ }$}} {\delte}}

\def\rmA{ {_{\rm \color{blue}A}}}  \def\alphab{ {\color{blue}\alpha} }

\def\thetar{ \colbrun{\Theta} } \def\varthetar{ \colbrun{\vartheta} }
\def\omegar{ \colbrun{\Omega} } \def\varpir{ \colbrun{\varpi} }
\def\varpirIJ{\spose {\raise 3.0pt\hbox{$\, \acute{\ }\grave{\ }$}}{\varpir}}

\def\calI{ {\color{red}{\cal I}} }    \def\calH{ {\color{red}{\cal H}} }
\def\calL{ {\color{red}{\cal L}} }    \def\Lred{ {\color{red} L} } 
\def\Cred{ {\color{red}C} }           \def\mred{ {\color{red}m} } 
\def\pred{ {\color{red}p} }           \def\Tred{ {\color{red}T} }
\def\pired{ {\color{red}\pi} } 

\def\ii{ {\color{vert}i} }            \def\ji{ {\color{vert}j} } 
\def\sigme{ {\color{vert}\sigma} }    \def\nable{ {\color{vert}\nabla} } 
\def\ete{ {\color{vert}\eta} }        \def\delte{ {\color{vert}\delta} }
\def\perpe{ {\color{vert}\perp} } 

\def\vq{ {\color{violet} q} }           
\def\vphi{ {\color{violet} \varphi} }

\def\bA{ {\colorange{A}} }

\def\bb{\  $ }  \def\fb{ $ }
\def\be{\begin{equation} } \def\fe{\end{equation}}
\begin{document}

\title{{\it \small 2002 Peyresq workshop contribution.}\\
 \textcolor{red}{\Large Symplectic 
Structure  in Brane Mechanics}    \\[1cm]
}

 \author{Brandon Carter \\
 \textcolor[named]{ForestGreen}{LuTh, Observatoire de Paris, 92 Meudon. } 
  }
\date{December, 2002}
\maketitle

\vskip 2 cm

\noindent
{\bf Abstract}
 
\medskip
This article treats the generalisation to brane dynamics
of the covariant canonical variational procedure leading to the construction
of a conserved bilinear symplectic current in the manner originally
developped by Witten, Zuckerman and others in the context of field
theory. After a general presentation, including a review of the
relationships between the various (Lagrangian, Eulerian and other) relevant
kinds of variation, the procedure is illustrated by application to
the particularly simple case of branes of the Dirac-Goto-Nambu type,
in which internal fields are absent.

\vfill\eject

\noindent
{\bf 1. Introduction}
\medskip

The purpose of this article to consider the application to classical brane
mechanics of the general principles of covariant canonical variational analysis, 
which provides a canonical symplectic structure, whose potential utility as a 
starting point for the covariant construction of corresponding quantum systems 
has been emphasised by Witten, Zuckerman, and 
others~\cite{Wi86,CrWi87,Zu87,So94,CF98,Nu00,Ro02} 
in  the context of relativistic field theories. The task of extending such 
analysis from ordinary fields to branes (meaning systems with support confined 
to a lower dimensional worldsheet) has recently been taken up by 
Cartas-Fuentevilla~\cite{CF02,CF02b}. The necessary analysis has been 
facilitated by the relatively new development~\cite{Ca93,BaCa95,BaCa00} of 
suitably covariant methods of geometrical analysis, which have already been 
shown to be far more efficient than the more cumbersome (and error prone) 
frame dependent methods used in earlier work for treating other problems, such 
as the divergences arising from self interaction~\cite{Ca97,CaBa98,CaBaUz02}.

One of the questions that has arisen in this work is that of how the conserved 
antisymmetric bilinear perturbation current that was obtained by a different 
approach in my original perturbation analysis~\cite{Ca93} of simple 
Dirac-Nambu-Goto type branes is related to the closed symmetric structure of 
the canonical treatment. The claim~\cite{CF02} that both approaches lead 
ultimately to the same result was based on an argument that, in its original 
version, depended on intermediate steps involving a questionable logical 
shortcut, but the ensuing conclusion is fully confirmed by the more rigorous 
and complete treatment provided here.

\bigskip\noindent
 {\bf 2. Brane variational principle}
\medskip

The present work will be concerned with the very broad category of
conservative p-brane models whose mechanical evolution is governed
by an action integral of the form
\be {\calI}=\int {\calL}\, {\rm d}^{p+1}\sigme\, ,\label{1}\fe
over a supporting worldsheet with internal co-ordinates  \bb\sigme^\ii\fb 
 \bb(\ii=0, 1, ... \,p )\, ,\fb and induced metric 
\bb {\ete}_{\ii\ji}=g_{\mu\nu}x^\mu_{,\ii}x^\nu_{,\ji}\fb,
in a background with coordinates 
\bb x^\mu\, ,\fb \bb(\mu=0, 1, ...\, d)\, ,\fb \bb (d\geq p)\ \fb
and (flat or curved) space-time metric \bb g_{\mu\nu}\, ,\fb.
The relevant Lagrangian scalar density is expressible in the form 
 \be{\calL}=\Vert{\ete}\Vert^{1/2}{\Lred}\, ,\label{2}\fe 
where \bb{\Lred}\fb is scalar function of a set of field  components 
\bb \vq^\rmA\ \fb -- including background coords and of their surface 
deriatives, \bb \vq^\rmA_{\, ,\ii}
=\partial_\ii\vq^\rmA=\partial\vq^\rmA/\partial\sigme^\ii\, .\fb
The relevant field variables  \bb\vq^\rmA\ \fb can be of internal or external 
kind, the most obvious example of the latter kind being the background
coordinates \bb x^\mu\fb themselves.

 The generic action variation,
 \be \delte{\calL}= {\calL}_{\!\rmA}\delte\vq^\rmA +
\pred_{\!\rmA}^{\, \ii}\delte\vq^\rmA_{\, ,\ii}\, ,\fe
specifies a set of partial derivative components \bb{\calL}_{\!\rmA} \fb 
and an  associated set of generalised momentum components
 \bb{\pred}_{\!\rmA}^{\, \ii}\fb. According to variation principle, the
dynamically admissible ``on shell'' configurations are those characterised
by the vanishing of the Eulerian derivative as given by 
\be \frac{\delte \calL}{\delte
\vq^\rmA}=\calL_{\!\rmA}-{\pred}_{\!\rmA\, ,\ii}^{\, \ii}\, .\fe

In terms of this Eulerian derivative, the generic Lagrangian variation 
will have the form
\be \delte\calL=\frac{\delte \calL}{\delte\vq^\rmA}
\delte\vq^\rmA+\big({\pred}_{\!\rmA}^{\, \ii}\delte\vq^\rmA
\big)_{,\ii}\, .\fe
There will be a corresponding pseudo-Hamiltonian scalar density
\be\calH= {\pred}_{\!\rmA}^{\, \ii}\vq^\rmA_{\, ,\ii}
-\calL\, ,\fe
for which
\be \delte\calH=\vq^\rmA _{\, ,\ii}\delte{\pred}_{\!\rmA}^{\, \ii}
-\calL_\rmA\delte\vq^\rmA      \, .\fe
(The covariance of such a pseudo Hamiltonian distingushes it from
the ordinary kind of Hamiltonian, which depends on the introduction
of some preferred time foliation.)

For an on-shell configuration, i.e. when the dynamical equations
 \be \frac{\delte \calL}{\delte\vq^\rmA}=0\, ,\fe
are satisfied, the Lagrangian variation will reduce to a pure surface 
divergence,
\be \delte\calL=\big({\pred}_{\!\rmA}^{\, \ii}\delte\vq^\rmA
\big)_{,\ii}\, ,\fe
and the correponding on-shell pseudo-Hamiltonian variation will take the form 
\be \delte\calH=\vq^\rmA _{\, ,\ii}\delte{\pred}_{\!\rmA}^{\, \ii}
-\pred_{\!\rmA\, ,\ii}^{\, \ii}\delte\vq^\rmA      \, .\fe

\bigskip
\noindent
{\bf {3. Canonical symplectic structure}}
\medskip

It is evident from the preceeding work that the generic first order
variation of the Lagrangian will be  expressible as 
\be \delte\calL=\frac{\delte \calL}{\delte\vq^\rmA}
\delte\vq^\rmA+\varthetar^\ii_{\, ,\ii}\, .\fe
in terms of the generalised Liouville 1-form 
(on the configuration space cotangent bundle) that is defined  by
\be \varthetar^\ii=\pred_\rmA^{\, \ii}\delte \vq^\rmA\, ,\fe

Let us now consider the effect of a pair of successive independent
variations \bb \deI\, ,\fb \bb \deJ\, , \fb
which will give a second order variation of the form 
\be \deJ\deI\calL=\deJ\Big(\frac{\delte \calL}{\delte\vq^\rmA}\Big)
\deI\vq^\rmA+\frac{\delte \calL}{\delte\vq^\rmA}\deJ\deI \vq^\rmA
+\big(\deJ\pred_\rmA^{\, \ii}\deI\vq^\rmA+
\pred_\rmA^{\, \ii}\deJ\deI\vq^\rmA\big)_{,\ii}\, .\fe
Thus using the commutation relation \bb \deJ\deI=\deI\deJ\fb one gets
\be
\deJ\Big(\frac{\delte \calL}{\delte\vq^\rmA}\Big)\deI\vq^\rmA-
\deI\Big(\frac{\delte \calL}{\delte\vq^\rmA}\Big)\deJ\vq^\rmA
=\varpirIJ^\ii_{\, ,\ii}\, ,\fe
where the symplectic 2-form (on the configuration space cotangent bundle) 
is defined by
\be \varpirIJ^\ii=
\deI\pred_\rmA^{\, \ii}\deJ\vq^\rmA-
\deJ\pred_\rmA^{\, \ii}\deI\vq^\rmA\, .\label{ome}\fe
 
For an on-shell perturbation we thus obtain
\be\frac{\delte \calL}{\delte\vq^\rmA}=0\hskip 1 cm \Rightarrow\ 
\hskip 1 cm \delte\calL=\varthetar^\ii_{\, ,\ii}\,,\fe
while for a pair of on-shell perturbations we obtain
\be\deI\Big(\frac{\delte \calL}{\delte\vq^\rmA}\Big)=
\deJ\Big(\frac{\delte \calL}{\delte\vq^\rmA}\Big)=0 \hskip 1cm
\ \Rightarrow\hskip 1cm \varpirIJ^\ii_{\, ,\ii}=0\, .\fe

The foregoing surface current conservation law is
expressible in shorthand notation as the condtion
\be\varpir^\ii_{\, ,\ii}=0\, ,\fe
 in which the closed (since manifestly exact) symplectic 2-form 
(\ref{ome}) is specified using concise wedge product notation as
\be \varpi^\ii=\delte\wedge\varthetar^\ii=\delte
\pred_\rmA^{\, \ii}\wedge\delte \vq^\rmA\, .\fe

It is to be remarked that some authors prefer to use an even more
concise notation system in which it is not just the relevant distinguishing
(in our case acute and grave accent) indices that are omitted but
even the wedge symbol \bb\wedge\fb that indicates the antisymmetrised
product relation. However such an extreme  level abbreviation is dangerous 
in contexts such as that of the present work in which symmetric products
are also involved, as is shown by the example~\cite{CF02} discussed
below, in which a formula involving a symmetric product was applied as if 
it were an antisymmetric product.

\bigskip\noindent
{\bf 4. Translation into strictly tensorial form.}
\medskip

In accordance with the strategy~\cite{Ca93} of avoiding the 
supplementary  gauge dependence involved in the use of
auxiliary structures such as local frames and internal surface
coordinates by working as far as
possible just with quantities that are strictly tensorial with respect
to the background space, it will be preferable for many purposes to
translate  the surface current densities whose components \bb 
\varthetar^\ii\fb and \bb \varpir^\ii\fb depend on the choice of 
the internal coordinates  \bb\sigme^\ii\fb,
into terms of the corresponding vectorial
quantities, which will have strictly tensorial background coordinate
 components given by
\be \thetar^\nu=\Vert\ete\Vert^{-1/2} x^\nu_{\, ,\ii}\varthetar^\ii
\, ,\hskip 1 cm
\omegar^\nu=\Vert\ete\Vert^{-1/2} x^\nu_{\, ,\ii}\varpir^\ii\, .\fe
These currents will have strictly scalar surface divergences given
 in terms of the corresponding scalar densities by
\be\overline\nable_{\!\nu}\thetar^\nu=\Vert\ete\Vert^{-1/2}
\varthetar^\ii_{\, ,\ii}\, ,\hskip 1 cm
\overline\nable_{\!\nu}\omegar^\nu=\Vert\ete\Vert^{-1/2}
\varpir^\ii_{\, ,\ii}\fe
where \bb\overline\nable\fb is the surface projected covariant differentiation
operator defined in terms of the fundamental tensor
\bb \ete^{\mu\nu}=\ete^{\ii\ji} x^\mu_{\, ,\ii}x^\nu_{\, ,\ji}\fb
by \bb\overline\nable_{\!\nu}=\ete^\mu_{\ \nu}\nabla_{\!\mu}\, .\fb

By the preceeding analysis, a Liouville current conservation law of the form
\be\overline\nable_{\!\nu}\thetar^\nu=0\fe
will hold for any symmetry generating perturbation, i.e. for any infinitesimal
variation \bb\delte\vq^\rmA\fb such that \bb\delta\calL=0\fb, and a symplectic
current conservation law of the form 
\be\overline\nable_{\!\nu}\omegar^\nu=0\fe
 will hold for any pair of perturbations that are on-shell, i.e.  
such that \bb\delte(\delte\calL/\delte q^\rmA)=0\fb.

\bigskip\noindent
{\bf{5. Covariant  variation formulae}}
\medskip

For physical evaluation of quantities the Liouville and symplectic currents 
\bb\thetar^\mu\fb and \bb\omegar^\mu\fb it is often more convenient to work 
with something less cordinate-gauge  dependent than the simple worldsheet 
based field component variations \bb \delte\vq^\rmA\fb used in preceding work.

In particular if the field component \bb\vq^\rmA\fb is of a kind that is
defined over the background -- not just confined to brane worldsheet with 
internal coordinates \bb\sigma^\ii\fb -- then with respect to a given system 
of external coordinates (which might, for example, be of Minkowski type if 
the background is flat) in terms of which \bb \partial_\ii \vq^\rmA
=x^\mu_{\, ,\ii}\partial_\mu \vq^\rmA\, ,\fb the field will have an 
\underline{Eulerian} (fixed background) variation \bb \dE \vq^\rmA\fb
that is well defined independently  of any choice of the internal
 coordinates \bb\sigme^\ii\fb, unlike the simple brane worldsheet variation, 
which will be given in terms of the relevant \underline{displacement vector}, 
\bb \xe^\mu=\delte x^\mu\, ,\fb
by
\be\delte\vq^\rmA= \dE \vq^\rmA +\xe^\mu\partial_\mu\vq^\rmA\, . \fe 

When one is dealing with a background field that is not simply a scalar
but of a more general tensorial nature, it will commonly be desirable to
go on to convert the \underline{Eulerian 
variation} formula
\be \dE =\delte -\vec\xe\cdot\partial \fe
into terms of \underline{covariant} derivation as given by
\be \vec\xe\cdot\!\nabla=\vec\xe\cdot\partial+
\{\vec\xe\cdot\Gamma\}\fe 
where \bb\{\vec\xe\cdot\Gamma\}\fb  is purely algebraic
operator involving contractions with 2-index quantity 
\bb(\vec\xe\cdot\Gamma)^\mu_{\ \nu}=
\xe^\rho\Gamma_{\!\rho\ \nu}^{\ \mu}\, ,\fb  as exemplified, 
for a vectorial (e.g. Killing) field \bb k^\mu\, ,\fb  or 
a covectorial (e.g. Maxwellian) form
\bb \bA_\mu\, ,\fb by
\be \{\vec\xe\cdot\Gamma\} k^\mu= (\vec\xi\cdot\Gamma)^\mu_{\ \nu}
k^\nu\, ,\hskip 0.8 cm \{\vec\xe\cdot\Gamma\} \bA_\mu= -
(\vec\xe\cdot\Gamma)^\nu_{\ \mu}\bA_\nu\, .\fe

Alternatively, instead of using the connection dependent covariant 
derivative, it may be more appropriate to work with the 
corresponding \underline{Lie derivative}, as  given by
a prescription of the form
\be \vec\xe\Libra=\vec\xe\cdot\nabla-\{\nabla\xe\} \, ,\label{lie}\fe
in which  the operator \bb\{\nabla\xe\}\fb acts by contractions
with the displacement gradient tensor \bb\nabla_{\!\nu}\xe^\mu,\fb 
in the manner exemplified respectively for
a vector \bb k^\mu\, ,\fb or  a 1-form (i.e. covector)
\bb \bA_\mu\, ,\fb by the formulae
\be \{\nabla\xe\} k^\mu= k^\nu\nabla_{\!\nu}\xe^\mu\, ,
\hskip 1 cm \{\nabla\xe\} \bA_\mu= -
\bA_\nu\nabla_{\!\mu}\xe^\nu\, .\fe
It can be seen that connection cancels out, so that the prescription
(\ref{lie}) will be equivalently 
expressible in terms just of  partial derivative components
\bb \partial_\nu\xe^\mu\fb as
\be \vec\xe\Libra=\vec\xe\cdot\partial-\{\partial\xe\}
\, .\fe 

Another kind of variation that is particularly important in the
context of brane mechanics -- because (unlike the Eulerian, covariant,
and Lie derivatives) it is always well defined even for fields whose
support is confined to the brane worldsheet -- is what is known as 
the \underline{Lagrangian variation}, meaning change with respect to
background coordinates that are dragged by displacement. In the case
of a field that is not confined to the brane worldsheet, so that
its Eulerian variation is well defined, this latter kind 
will be related to the corresponding Lagrangian variation
by the well known Lie derivation formula
\be\dL=\dE+\vec\xe\Libra\, .\label{LagEul}\fe 

Yet another possibility that may be useful is to express the
 \underline{Eulerian} 
(fixed background point) variation in the form
\be \dE=\dP -\vec\xe\cdot\!\nabla\, ,\fe
where \underline{parallely transported variation} defined -- not just 
for background field, but also for tensor confined to brane -- by 
\be \dP=\delte + \{\vec\xe\cdot\Gamma\}\, ,\fe
using operator notation introduced above.

Unlike the covariant and Lie derivations \bb\vec\xe\cdot\!\nabla\fb
and \bb\vec\xe\Libra\fb and unlike the
Eulerian variation \bb\dE\fb, the \underline{parallel} variation 
\bb\dP\fb shares with the \underline{Lagrangian} variation \bb\dL\fb the 
important property of being well defined not just for background fields
but also for fields whose support is confined to the brane worldsheet.
The \underline{Lagrangian} variation \bb\dL\fb will always be expressible
directly in terms of the corresponding \underline{parallel} variation 
\bb\dP\fb by a relation of the form
\be \dL=\dP - \{\nabla\vec\xe\}\, ,\fe
in which it can be seen that connection dependence cancels out, 
leaving an expression of the simple form \bb\{\nabla\vec\xe\}\fb
\be \dL=\delte - \{\partial\vec\xe\}\, ,\fe
where the action of the algebraic operator \bb \{\partial\vec\xe\}\fb
is exemplified for a vector \bb k^\mu\fb, or a covector
\bb \bA_\mu\fb, by the respective formular
\be \{\partial\vec\xe\} k^\mu= k^\nu\partial_{\nu}\xe^\mu\, ,
\hskip 1 cm \{\partial\vec\xe\} \bA_\mu= -
\bA_\nu\partial_{\!\mu}\xe^\nu\, .\fe

In conclusion of this overview of the relationships between the various
kinds of infinitesimal variations that are commonly useful, it is to
be mentionned that in literature dealing with purely non relativistic 
contexts in which it is possible (though not necessarily wise) to
work exclusively with space coordinates of strictly Cartesian (orthonormal)
type, the variations of the kind referred to here as ``parallel'' are
generally described as ``Lagrangian'' by many authors.  That usage 
does not necessarily lead to confusion, because for scalars the distinction
does not arise, and because such authors systematically eschew the use 
(and the technical advantages) of Lagrangian variations of the
fully comoving kind (that is considered here) by working exclusively
with tensor components that are evaluated in terms only of orthonormal frames.

\bigskip\noindent 
{\bf 6. Evaluation in terms of Lagrangian variations.}
\medskip

In typical applications, the relevant set of configuration
 components \bb \vq^\rmA\fb 
will include a set of brane field components \bb \vphi^\alphab\fb 
as well as the background coords \bb x^\mu\fb, so that in terms of 
displacement vector
\bb\xe^\mu=\delte x^\mu\fb the
Liouville current will take the form
\be \thetar^\nu=\Vert\ete\Vert^{-1/2} x^\nu_{\, ,\ii}\big(
\pred_\alphab^{\ii}\,\delte\vphi^\alphab+ 
\pred_\mu^{\ \ii}\,\xe^\mu\big)=
\pired_\alphab^{\nu}\,\delte \vphi^\alphab+
\pired_\mu^{\ \nu}\,\xe^\mu\, ,\fe 
in which the latter version replaces the original momentum components
by corresponding background tensorial momentum variables that are
defined by
\be \pired_\alphab^{\,\nu}=\Vert\ete\Vert^{-1/2}\, x^\nu_{\, ,\ii}\,
\pred_\alphab^{\ii}\fe 
and
\be \pired_{\!\mu}^{\ \nu}=\Vert\ete\Vert^{-1/2}\, x^\nu_{\, ,\ii}\,
\pred_\mu^{\ \ii }\fe.

In order to obtain an analogously tensorial formula for the symplectic 
current 2-form, it is convenient, as a first step, to take advantage of 
the symmetry property
  \bb \Gamma_{\!\mu\ \rho}^{\ \nu} =
\Gamma_{\!\rho\ \mu}^{\ \nu},\fb of the Riemannian connection of the
background spacetime metric, which allows substitution of 
 \underline{parallel} variation  
\bb \dP\pred_\mu^{\ \ii}=\delte\pred_\mu^{\ \ii}
- \Gamma_{\!\mu\ \rho}^{\ \nu}\pred_\nu^{\ \ii}\xe^\rho\fb
for \bb \delte\pred_\mu^{\ \ii}\fb so as to provide an expression of 
the form
\be \omegar^\nu=\Vert\ete\Vert^{-1/2} x^\nu_{\, ,\ii}\big(
\delte \pred_\alphab^{\ii}\wedge\delte\vphi^\alphab+
\dP\pred_\mu^{\ \ii}\wedge\xe^\mu\big)\, .\fe
The next step is to evaluate the relevant momentum variations in
terms of the corresponding \underline{Lagrangian} variations,
using the formulae
\be \Vert\ete\Vert^{-1/2} x^\nu_{\, ,\ii}\,\delte
\pred_\alphab^{\ii} =\dL\pired_\alphab^{\,\nu}+\pired_\alphab^{\,\nu}
\overline\nable_{\!\rho}\xe^\rho\, ,\fe 
and
\be \Vert\ete\Vert^{-1/2} x^\nu_{\, ,\ii}\,\dP
\pred_\mu^{\ \ii }=\dL\pired_{\!\mu}^{\ \nu}-\pired_{\!\rho}^{\ \nu}
\nabla_{\!\mu}\xe^\rho+\pired_{\!\mu}^{\ \nu}\overline\nable_{\!\rho}
\xe^\rho\, . \fe 

The advantage of \underline{Lagrangian variations} is their convenience
for relating the relevant intrinsic physical quantities via the
appropriate equations of state.
 
\bigskip\noindent
{\bf 7. The simply elastic category}
\medskip

The illustration that follows will be restricted to the
\underline{simply elastic} category (including the case of an ordinary 
barotropic perfect fluid) in which -- with respect to a suitably 
comoving internal reference system \bb \sigme^\ii\fb -- there are no 
independent surface fields at all --  meaning that the
\bb \vphi^\alphab\fb and the \bb\pred_\alphab^{\,\ii}\fb are absent --
and in which the only relevant background field is the metric 
\bb g_{\mu\nu}\fb that is specified as a function of the external
coordinates \bb x^\mu\fb.

In any such simply elastic case, the generic variation of the Lagrangian is 
fully determined by the relevant surface stress momentum energy density tensor 
\bb\overline\Tred{^{\mu\nu}}\fb 
according to the standard prescription 
\bb \delte\calL=\frac{1}{2}\Vert\ete\Vert^{1/2}\,
\overline\Tred^{\mu\nu}\,\dL g_{\mu\nu}\, ,\fb 
whereby \bb\overline\Tred{^{\mu\nu}}\fb is specified in terms of partial 
derivation of the action density with respect to the metric.  In a fixed 
background (i.e. in the absence of any Eulerian variation of the metric) 
the Lagrangian variation of the metric will be given, according to the 
formula (\ref{LagEul}), by
\bb \dL g_{\mu\nu}=\vec\xe\Libra g_{\mu\nu}
=2\nabla_{\!(\mu}\xe_{\nu)}\, .\fb
By omparing this to  canonical prescription 
\bb\delte\calL=\calL_\mu\xe^\mu
+\pred_\mu^{\ \ii}\xe^\mu_{\, ,\ii}\fb 
with 
\bb\xe^\mu=\delte x^\mu\fb
it can be seen that the relevant partial derivatives will be given by
the (non-tensorial) formulae
\bb\calL_\mu=\Vert\ete\Vert^{1/2}\,\Gamma_{\!\mu\ \rho}^{\ \nu}
\overline\Tred_{\!\nu}{^\rho}\fb 
and 
\bb \pred_\mu^{\ \ii}=
\Vert\ete\Vert^{1/2}\,\overline \Tred_{\mu\nu}
\ete^{\ii\ji} x^\nu_{\, ,\ji}\fb.

The next step is to translate the result into background tensorial form.
It can be seen from the preceding work that in the \underline{simply elastic} 
case, the canonical momentum tensor \bb \pired_\mu{^{\ \nu}}\fb 
and the Liouville current \bb\thetar^\nu\fb
will be given just in terms of surface stress tensor
\bb\overline{\Tred}{^{\mu\nu}}\fb by the very simple formulae
\be \pired_\mu{^{\nu}}=\overline{\Tred}_\mu{^{\nu}}\, ,\hskip 1 cm
\thetar^\nu=\overline{\Tred}_\mu{^{\nu}}\xe^\mu\, .\fe

In order to proceed, we must consider the second order metric 
variation, whereby (following Friedman and Schutz~\cite{FrSc75})
the hyper Cauchy tensor (generalised elasticity tensor) 
\bb \overline{\Cred}{^{\mu\nu\rho\sigma}}
=\overline{\Cred}{^{\rho\sigma\mu\nu}}\fb
is specified~\cite{BaCa95} in terms of Lagrangian variations by
a partial derivative relation of the form
\be\dL\big(\Vert\ete\Vert^{1/2}\, \overline{\Tred}{^{\mu\nu}}\big)=
\Vert\ete\Vert^{1/2}\overline{\Cred}{^{\mu\nu\rho\sigma}}
\dL g_{\rho\sigma}\fe. 
The symplectic current is thereby obtained in the form
\be \omegar^\nu=\big(2\overline{\Cred}{_{\mu\ \rho}^{\ \, \nu\ \sigma}}
\overline\nable_{\!\sigme}\xe^\rho+\overline{\Tred}{^{\nu\rho}}
\overline\nable_{\!\rho}\xe_\mu\Big)\wedge \xe^\mu\, . \label{elas} \fe

\bigskip\noindent
{\bf 8. The simple the Dirac Goto Nambu case}
\medskip

The perfectly elastic category to which the formula (\ref{elas}) is 
applicable includes examples such as the case (to which much attention 
has been given in recent work on cosmology) of 3-brane world model
with a matter content consisting of a barotropic perfect fluid matter.

The consideration of such cases will however be left for future work,
while the present article will be concluded by the treatment of the
relatively trivial special case of a Dirac-Goto-Nambu type brane,
i.e. a brane on which their are no internal fields at all, so that
the Lagrangian scalar \bb\Lred\fb introduced in (\ref{2}) will simply
be a constant, which will be expressible in the form
\be \Lred= -\mred^{p+1}\label{Drac}\fe 
for some fixed mass scale \bb\mred\fb.

In terms of the of tangential and orthogonal projectors \bb\ete^\mu_{\ \nu}\fb 
and \bb\perpe^{\!\mu}_{\,\nu}=g^\mu_{\ \nu}-\ete^\mu_{\ \nu}\fb, it can be 
seen that for the Dirac-Nambu-Goto case characterised by (\ref{Drac})
the surface stress energy momentum density tensor will be given by an
expression of the familiar simple form
\be \overline\Tred^{\mu\nu}= -\mred^{p+1}\ete^{\mu\nu}\fe 
while the generalised Cauchy tensor will be obtained~\cite{BaCa95} in
the (less well known) form
\be \overline{\Cred}{^{\rho\sigma\mu\nu}}=
\mred^{p+1}(\ete^{\mu(\rho}\ete^{\sigma)\nu}-\frac{1}{2}\ete^{\mu\nu}
\ete^{\rho\sigma})\, .\fe 
It can thus be seen that the canonical symplectic current (\ref{elas}) will 
be given explicitly by the formula
\be\omegar^\nu=\mred^{p+1}\big(\ete^{\nu\sigma}\perpe_{\mu\rho} +
2\ete_\mu^{\, [\nu}\ete_\rho^{\,\sigma]}
\big)\xe^\mu\wedge\overline\nable_{\!\sigma}\xe^\rho\, .\fe

It can now be checked by direct comparison that this canonical symplectic
current does indeed agree with the antisymmetric bilinear current that I 
originally obtained~\cite{Ca93} by a rather different approach. The claim
by Cartas-Fuentaevilla that this bilinear current is of canonical (closed
since exact) type is thereby confirmed. 

The reason why the original argument 
to this effect~\cite{CF02} was not entirely convincing was that it depended on
an assumption to the effect that \bb\xe^\nu\wedge\nabla_{\!\nu}\xe^\mu\fb
should vanish. The meaning of this condition is that the pair of 
displacement vector fields involved should commute, something that could 
always  be imposed for a brane in the restricted sense (but not for not for 
the dimensionally maximal limit case of a space filling fluid or solid 
medium) by using the gauge freedom to make arbitrary adjusments of the choice 
of the displacement field off the world sheet where it has no physical 
effect. However instead of being invoked as a (perfectly legitimate) choice 
of gauge, the commutator was expressed using the dangerously ambiguous 
abbreviation scheme in which the wedge symbol \bb\wedge\fb was omitted so 
that it took the form \bb\xe^\nu\nabla_{\!\nu}\xe^\mu\fb, whose vanishing was 
accounted for on the basis of a reinterpretation as if the product were of 
symmetric type, involving just a single displacement vector field 
\bb \xe^\nu\fb, which was thereby required to be geodesic. It happens that 
this (unnecessary and insufficient) condition of geodicity could also 
(if genuinely needed) be imposed (on one but not both of the commuting vector 
fields) as a choice of gauge off the worldsheet, but it was  unjustifiably 
alleged~\cite{CF02} to be implicit as a necessity for my method of 
analysis~\cite{Ca93}. 

Despite of the fact that it does not have to apply in general, the 
consideration that the litigious intermediate requirement (namely
the simplification provided by the vector commutation condition, not to 
mention the quite redundant geodicity condition) that was invoked~\cite{CF02} 
can actually be imposed as an admissible choice of gauge, means that if used 
more carefully it could after all provide a logically valid chain of 
reasonning leading to the final (gauge invariant) conclusion -- albeit by a 
route that is less explicit and direct than that of the present article 
(which makes no use of any gauge restrictions at all).

To complete this clarification, I would  emphasise that my method does 
not depend on any (geodesic or other) restriction on the choice of the 
displacement vector field off the world sheet. (This means that the method 
is applicable, not just to branes in the restricted sense, but also to 
ordinary space filling solids and fluids, for which there cannot be any
freedom to adjust the displacement field, because no off shell region is 
available.) As discussed in more detail in a more recent review~\cite{Ca96}, 
my system of analysis does indeed involve the use of geodicity:  however 
it is not invoked as as a restriction on the infinitesimal displacement 
field \bb\xe^\nu\fb but merely as a means of using an (entirely arbitrary) 
infinitesimal displacement field to specify a corresponding finite 
displacement.

\bigskip\noindent
{\bf Acknowledgements.}
\smallskip

I wish to thank D. Steer for technical discussions, and to thank E. 
Gunzig and E. Verdaguer for organising the Peyresq meeting that was 
the occasion for undertaking this work.

\end{document}